\documentclass[11pt]{article}
\newcommand{\Comment}[1]{{}}
\usepackage{amsfonts,amsthm,amsmath,amssymb,slashed}
\usepackage[textwidth = 430 pt, textheight = 630 pt]{geometry}
\usepackage{color}

\Comment{\usepackage{color}
\definecolor{MyDarkBlue}{rgb}{0.15,0.15,0.45}
\usepackage[linktocpage=true]{hyperref}
\hypersetup{
colorlinks=true,
citecolor=MyDarkBlue,
linkcolor=MyDarkBlue,
urlcolor=MyDarkBlue,
pdfauthor={Horatiu Nastase},
pdftitle={General $f($) and conformal inflation from minimal supergravity plus matter},
pdfsubject={hep-th}
}

\usepackage[numbers,sort&compress]{natbib}
\usepackage{hypernat}}
\usepackage{graphicx}
\usepackage{cite}

\newcommand\ignore[1]{}
\def\one{{\,\hbox{1\kern-.8mm l}}}

\def\a{\alpha}\def\b{\beta}

\def\d{\partial}

\newcommand{\Cset}{{\,\,{{{^{_{\pmb{\mid}}}}\kern-.45em{\mathrm C}}}}}

\newcommand{\be}{\begin{equation}}
\newcommand{\bea}{\begin{eqnarray}}

\newcommand{\ee}{\end{equation}}
\newcommand{\eea}{\end{eqnarray}}

\providecommand{\gsim}{\gtrsim}

\begin{document}

\renewcommand{\thefootnote}{\fnsymbol{footnote}}

\makeatletter
\@addtoreset{equation}{section}
\makeatother
\renewcommand{\theequation}{\thesection.\arabic{equation}}

\rightline{}
\rightline{}


\vspace{10pt}


\begin{center}
{\LARGE \bf{\sc Small field inflation in ${\cal N}=1$ supergravity with a single chiral superfield}}
\end{center} 
 \vspace{1truecm}
\thispagestyle{empty} \centerline{
{\large \bf {\sc Heliudson Bernardo}}\footnote{E-mail address: \Comment{\href{mailto:heliudson@ift.unesp.br}}{\tt heliudson@ift.unesp.br}}
     {\bf{\sc and}}
{\large \bf {\sc Horatiu Nastase}}\footnote{E-mail address: \Comment{\href{mailto:nastase@ift.unesp.br}}{\tt nastase@ift.unesp.br}}
                                             }

\vspace{.5cm}


\centerline{{\it 
Instituto de F\'{i}sica Te\'{o}rica, UNESP-Universidade Estadual Paulista}} \centerline{{\it 
R. Dr. Bento T. Ferraz 271, Bl. II, Sao Paulo 01140-070, SP, Brazil}}

\vspace{1truecm}

\thispagestyle{empty}

\centerline{\sc Abstract}

\vspace{.4truecm}

\begin{center}
\begin{minipage}[c]{380pt}
{\noindent We consider "new inflation" inflationary models at small fields, embedded in minimal ${\cal N}=1$ supergravity with a single chiral superfield. 
Imposing a period of inflation compatible with experiment severely restricts possible models, classified in perturbation theory. 
If moreover we impose that the field goes to large values
and very small potential at the current time, like would be needed for instance for the inflaton being the volume modulus in large extra dimensional
scenarios, the possible models are restricted to very contrived superpotentials. 
}
\end{minipage}
\end{center}

\vspace{.5cm}

\setcounter{page}{0}
\setcounter{tocdepth}{2}

\newpage

\renewcommand{\thefootnote}{\arabic{footnote}}
\setcounter{footnote}{0}

\linespread{1.1}
\parskip 4pt



\section{Introduction}
\ \ \ \ \
Inflation is now the standard paradigm for the beginning of the Universe, though there are alternatives, like ekpyrotic and cyclic models 
\cite{Khoury:2001wf,Buchbinder:2007ad,Steinhardt:2001st}, 
string and brane gas cosmologies \cite{Brandenberger:1988aj,Nayeri:2005ck,Alexander:2000xv}, 
etc., but none are as compelling. On the other hand, supersymmetry is a standard tool in particle physics
to fix a variety of problems, like the hierarchy problem, gauge coupling unification, offering natural dark matter candidates, and generally improving
the UV behaviour of the Standard Model. Despite the (yet) lack of experimenal evidence, being such a compelling scenario, it is still the 
preferred model for an extension of the Standard Model. When considering a model that includes gravity, like in the case of inflationary 
cosmology, we have to consider supergravity, and the simplest case is to consider ${\cal N}=1$ minimal supergravity plus matter. 

It would be the natural thing to ask if we can embed the inflaton into a chiral superfield that couples to the minimal supergravity. 
It turns out that having {\em just} the inflaton chiral superfield coupled to supergravity is quite difficult to achieve (see e.g. the 
review \cite{Yamaguchi:2011kg}).  Progress towards embedding a general inflationary model in minimal supergravity plus a single 
chiral superfield was achieved in \cite{Ketov:2014qha,Ketov:2014hya}, with a K\"{a}hler potential that is somewhat non standard.\footnote{Also, a model 
of helical inflation in minimal supergravity with a single  chiral superfield was obtained in \cite{Ketov:2015tpa}. Within the context of cosmological 
$\a$-attractors a model of minimal supergravity with a single chiral superfield was obtained in \cite{Roest:2015qya}.}
The case of generalized $f(R)$ inflation was analyzed in \cite{Nastase:2015pua}. 

But one of the most popular ways to obtain inflation is in the "new inflation" scenario, where there is a plateau at small inflaton field, followed by 
a drop in the potential, that realizes reheating or preheating. It is therefore of interest to embed it in minimal supergravity plus a single chiral 
superfield, but that is in fact quite difficult. It is generally considered that the K\"{a}hler potential at small field has either a perturbative form 
around the canonical form, or a logarithmic form that is obtained for instance if the inflaton is a volume modulus. For the (nonperturbative)
superpotential, one can consider the most general form possible.

Therefore, in this paper we will investigate the generality of obtaining "new inflation" within the context of minimal supergravity coupled
with a single scalar superfield. We will consider the various possibilities for the K\"{a}hler potential and superpotential, viewed as a 
perturbative expansion around a leading term. We will see that this puts stringent constraints on the possible forms of the superpotential. 
Further, we will investigate what happens if we want also the scalar to be able to reach large values at small potential (small cosmological 
constant), like in the case the inflaton is the volume modulus in large extra dimensions. We will see that then only very contrived forms of the 
superpotential are allowed, and the experimental constraints further rule out most models. 

The paper is organized as follows. In section 2 we analyze small field inflation.\footnote{Small field inflation in ${\cal N}=1$ supergravity 
was also briefly analyzed  in \cite{Terada:2015sna}.} After a general description of the possibilities allowed by theory 
and experiment, classifying the possible K\"{a}hler potentials and superpotentials, we show the effect of field redefinitions in section 2.1. 
After that, we proceed to analyze the possible cases of K\"{a}hler potentials and superpotentials in section 2.2. In section 2.3 we consider 
the constraints imposed by larger fields, in order to {\em end} inflation, in a reheating or preheating phase. 
Then in section 3 we consider the constraints imposed if we want to be able to reach high values of the scalar field, for small values of the potential. 
We conclude in section 4.

\section{Small field inflation}

Consider that the inflaton $\sigma$ is the imaginary part of a complex scalar $\Phi$ that is part of a chiral supermultiplet, coupled to minimal supergravity, 
i.e. $\Phi=i\sigma+...$ The choice of real or imaginary part is a convention on whether we multiply $\Phi$ with an $i$.  

We want to make an analysis of the simplest and the most natural possibilities for the K\"{a}hler potential $K(\Phi,\bar\Phi)$ and the 
superpotential $W(\Phi)$, and impose the usual constraints on the resulting potentials.\footnote{Some constraints on inflationary models coming  from 
K\"{a}hler geometry were found in \cite{Covi:2008cn,Hetz:2016ics}.}

First off, $K(\Phi,\bar\Phi)$ must be a real function, because of symmetry (though one could imagine a more 
general function of $\Phi$ and $\bar \Phi$ separately, and then in the $x$-space action one would add the hermitian conjugate, but 
in this way nothing new would be obtained). Then, we can think of it as being a function of either $\bar \Phi \Phi$ or $\Phi+ \bar\Phi$.
Note that we could have also $i(\bar\Phi-\Phi)$, but that is simply related by a rescaling with an $i$ from the $\Phi+\bar\Phi$ case. 
However, if we define that the inflaton is the {\em imaginary } part of $\Phi$, then we should add both kinds of corrections.

Given a K\"{a}hler potential $K$ and a superpotential $W$, the scalar potential is given by the formula
\be
V= e^{K/M_{\rm Pl}^2}\left[g^{i\bar j}D_i W\overline{D_j W}-3\frac{|W|^2}{M_{\rm Pl}^2}\right]\;,
\ee
where $g_{i\bar j}=\d_i\d_{\bar j}K$ as before and
\be
D_iW=\d_i W+\frac{1}{M_{\rm Pl}^2}(\d_iK)W.
\ee

Unless explicitly stated, we will put $M_{\rm Pl}=1$ in the following. 

In a perturbative expansion, the K\"{a}hler potential could:

\begin{itemize}

\item - start off as the canonical potential $\bar \Phi \Phi$, that gives a canonical kinetic term for the scalars in the superfield, and 
then have small corrections (when $\sigma={\rm Im}\Phi$ is small) around it. Thus
\be
K(\bar\Phi,\Phi)=\bar \Phi \Phi+{\cal O}((\bar \Phi\Phi)^2)+{\cal O}((\bar \Phi+\Phi)^2)+{\cal O}([i(\bar\Phi-\Phi)]^2).
\ee
In this way, $K$ has a Taylor expansion (in the complex variables).

\item -start off as a real linear term, plus corrections, i.e. 
\be
K(\bar\Phi,\Phi)=\bar \Phi+ \Phi+{\cal O}(\bar \Phi\Phi)+{\cal O}(\bar \Phi+\Phi)+{\cal O}(i(\bar\Phi-\Phi)).
\ee
Here the leading term gives a vanishing kinetic term, so in fact we need to consider the quadratic part as leading, and then we will 
see that this contains an important case. 

\item -start off as a singular function with only poles, i.e. 
\be
K(\bar\Phi,\Phi)=\frac{1}{(\bar \Phi \Phi)^n}\left[1+{\cal O}(\bar \Phi\Phi)+{\cal O}(\bar \Phi+\Phi)+{\cal O}(i(\bar\Phi-\Phi))\right].
\ee
It is not very clear how could the leading term appear from a more fundamental origin like string theory, 
as there are no known examples for its origin.

\item -start off as a singular function with an essential singularity. There are no known examples for something like $e^{-\frac{1}{\bar\Phi\Phi}}$
for the leading term, which would be a natural guess for an essential singularity. But there are known examples for a logarithmic form
as the leading term, which we will write more generally as 
\be
K(\bar \Phi,\Phi)=\a\ln\left(a(\bar\Phi+\Phi)+bi(\bar \Phi-\Phi)\right)\left[1+{\cal O}(\bar \Phi\Phi)
+{\cal O}(\bar \Phi+\Phi)+{\cal O}(i(\bar\Phi-\Phi))\right]\;,\label{logk}
\ee
where $a$ and $b$ are real, and we put both $a$ and $b$ terms since we fixed the inflaton to be {\em imaginary} part of $\Phi$, but both 
real and imaginary parts could appear in principle inside the log. This is in fact a very important example, since for instance the K\"{a}hler 
potential for the volume modulus in supergravity compactifications has this logarithmic form. Also other moduli in string compactifications
can appear inside the log. If then one of these moduli is the inflaton, we naturally obtain the case above. 

\end{itemize}

For the superpotential $W(\Phi)$, we can also consider a similar classification, 
in terms of a leading term that is either a positive power law
$\Phi^n$, a negative power law (pole) $\Phi^{-m}$, or an essential singularity, which naturally could be $e^{-\frac{1}{\Phi^n}}$ or 
$\ln \Phi$. Note that now we have a single variable, the holomorphic variable $\Phi$, so the leading term is simply multiplied with a 
generic Taylor expansion in $\Phi$, $\sum_{n\geq 0}c_n\Phi^n$. 

Before we start this analysis however, we will say a few words about possible field and function redefinitions.

\subsection{Field redefinitions}

There are various field redefinitions one could make. First of all, for complex scalars $\phi^i$ with a generic sigma model kinetic term, 
\be
-\frac{1}{2}\int d^4x g_{i\bar j}(\phi^k)\d^\mu\phi^i\d_\mu\bar\phi^{\bar j}\;,
\ee
which in the supersymmetric context comes from the K\"{a}hler potential for the superfields $\Phi^i=\phi^i+{\cal O}(\theta)$,
\be
g_{i\bar j}=\d_i\d_{\bar j}K(\bar\phi^{\bar i},\phi^j)\;,
\ee
we can always make arbitrary scalar field redefinitions $\phi'^i=f(\bar \phi^{\bar k},\phi^l)$ to change the kinetic term to whatever 
we want, for instance to put it
to the canonical form (with $g_{i\bar j}=\delta_{i\bar j}$).\footnote{Of course, this is true only locally, globally there could be obstructions. Since 
however here we are interested mostly in models of small field inflation, that is not important.} 
However, if we consider instead only the restricted set of {\em holomorphic} 
transformation, 
\be
\phi^i=f(\phi^k)\;,
\ee
which can be extended to the holomorphic transformations of the superpotential, which therefore preserve the manifestly 
supersymmetric structure,
\be
\Phi^i=f(\Phi^k)\;,
\ee
then we cannot guarantee that we can put the sigma model metric in whatever form we want, in particular we cannot guarantee that we 
can obtain the canonical form (with $g_{i\bar j}=\delta_{i\bar j}$). 

In addition to these transformations on the fields, there is another ambiguity that only affects the K\"{a}hler potential $K$, namely the 
K\"{a}hler transformations, which are a sort of "gauge" transformations of $K$ that leave the metric $g_{i\bar j}$ invariant, and thus 
the form of the explicit Lagrangean of rigid supersymmetry; in supergravity we must supplement them (in the case when $f_1=f_2\equiv f$ below)
with a change in the superpotential in order to have the bosonic Lagrangian be invariant. They are given by
\be
K\rightarrow K+f_1(\Phi)+f_2(\bar \Phi)\;,
\ee
as well as $W\rightarrow e^{-f} W\equiv W'$. In the following we will consider implicitly that we work with $W'$. 

We want to see what is the effect of the K\"{a}hler transformations and the holomorphic redefinitions on the perturbative $K$, i.e. what can we fix 
by using these transformations. Note that if we want $K$ to remain real, we need $f_1=f_2$.

\subsubsection{Leading canonical term in $K$}

Consider a K\"{a}hler potential that starts with the canonical term $\bar\Phi \Phi$, and then includes higher order corrections. For a generic one, 
we expect to have the possible corrections enumerated in the previous subsection. However, we can consider also the special case when 
the next {\em nonzero} correction occurs only at order $n$ in $\Phi,\bar\Phi$. 

Then in some special cases, we can get rid of the corrections as follows.

\begin{itemize}

\item If $K = \Phi\bar{\Phi} + \a(\Phi^n + \bar{\Phi}^n)$, the K\"{a}hler potential is equivalent to the canonical one by the K\"{a}hler transformation
\be
K \longrightarrow K - \a\Phi^n - \a \bar{\Phi}^n.
\ee			
\item If $K = \Phi\bar{\Phi} + \gamma(\Phi + \bar{\Phi})^n$, we can instead use the field redefinition $\Phi \longrightarrow \Phi - \sum_k
\beta_k \Phi^k$ to write
\bea
\Phi\bar{\Phi} &=& \Phi'\bar{\Phi}' + \sum_k\beta_k(\Phi^k\bar{\Phi}' + \bar{\Phi}^k\Phi') + \sum_{k,l}\beta_k\beta_l \bar{\Phi}^k\Phi^l\nonumber\\
&&\simeq \Phi'\bar{\Phi}' + \sum_k\beta_k(\Phi^k\bar{\Phi} + \bar{\Phi}^k\Phi)\;,
\eea	
\end{itemize}
where in the last line we have considered that $\Phi$ is small. 

In general, at order $n$ we could have corrections of the type $\Phi^n,\Phi^{n-1}\bar\Phi,\Phi^{n-2}\bar\Phi^2,...\bar\Phi^n$, of course 
with the constraint of reality, which would relate the coefficients. 
The redefinition and the K\"{a}hler transformation above can be used to cancel the terms $\Phi^n$, $\bar{\Phi}^n$ and 
$n(\Phi^{n-1} \bar{\Phi} +\bar{\Phi}^{n-1} \Phi)$, but that would leave (if $n\neq 2$, i.e. if we have a {\em correction} to the canonical 
case, which itself has $n=2$) other terms unchanged, and in this case there will be more than one term.

As an example, consider corrections that are only functions of $\bar\Phi+\Phi$, and consider their expansion up to order $n=4$. 
Then we have generically
\bea
K&=& \bar{\Phi}\Phi + \alpha(\Phi + \bar{\Phi})^3 + \beta(\Phi +\bar{\Phi})^4\cr
&= &\bar{\Phi}\Phi + \alpha (\Phi^3 +\bar{\Phi}^3) + 3\alpha(\Phi^2\bar{\Phi} + \bar{\Phi}^2\Phi) + \beta(\Phi^4+\bar{\Phi}^4)+ \cr
&&+ 4\beta(\Phi^3\bar{\Phi} +\Phi\bar{\Phi}^3) +6 \beta(\Phi^2\bar{\Phi}^2).
\eea 
We see then that considering a field redefinition with $\beta_2 = -3\alpha$, $\beta_3 = -4\beta$ and  a K\"{a}hler transformation with 
$F(\Phi) = -\alpha\Phi^3 -\beta\Phi^4$, we can get rid of all terms except the symmetric one, $\bar\Phi^2\Phi^2$, so the $\a$ and $\b$ 
corrections in this example are equivalent to a $\bar\Phi^2\Phi^2$ correction.

\subsubsection{Leading logarithmic term in $K$}

Another useful possibility is to have a logarithmic K\"{a}hler potential, with corrections. For instance, if $\Phi$ is a volume modulus
for the compact space $K$, specifically $\sigma={\rm Im}\Phi$ being the volume in Planck units up to some numbers, then 
from the Kaluza-Klein dimensional reduction on $K$ we obtain (reinstating $M_{\rm Pl}$ momentarily for emphasis)
\be
\frac{K}{M_{\rm Pl}^2}=-\a\ln\left[\frac{-i(\Phi-\bar\Phi)}{M_{\rm Pl}}\right]\;,
\ee
with $\a=3$ for a 6-dimensional space $K_6$, and in general $\a$ depending only on the dimension. Moreover, in this case 
the corrections naturally appear inside the log, as exemplified for instance by string corrections in various string models. 
Consider therefore
\be
\frac{K}{M_{\rm Pl}^2}=-\a\ln\left[\frac{-i(\Phi-\bar\Phi)+\kappa f(\Phi,\bar\Phi)}{M_{\rm Pl}}\right].
\ee
We can in principle consider both the case when $f$ is leading, or subleading, at small $\Phi$. At this time we will consider only the case it is 
subleading, coming back to the leading case in the next section. 

Then a field redefinition at small field $\Phi$, i.e. $\Phi'=\Phi+\b_k\Phi^k$ would simply redefine the function $f$. On the other hand, a 
K\"{a}hler transformation would simply turn it into the more general
\be
\frac{K}{M_{\rm Pl}^2}=-\a\ln\left[\frac{-i(\Phi-\bar\Phi)+\kappa f(\Phi,\bar\Phi)}{M_{\rm Pl}}\right]+f_1(\Phi)+f_2(\bar \Phi).
\ee

In this case, the scalar potential (for a general superpotential $W$) will be 
\bea
V&=& \left(-i\frac{\phi - \bar{\phi} + i \kappa f}{M_{\rm Pl}}\right)^{-\alpha}\left\lbrace\left[\frac{(\phi - \bar{\phi} +i\kappa f)^2/M_{\rm Pl}^2}
{-\alpha i\kappa(\phi - \bar{\phi} 
+i\kappa f)\partial_{\phi}\partial_{\bar{\phi}}f - \alpha(1+i\kappa \partial_{\phi}f)} \right] \times \right.\cr
&&\left.\times\left[\partial_{\phi}W - \frac{\alpha W(1 +i \kappa \partial_{\phi}f)}{\phi - \bar{\phi} + i\kappa f} \right]\left[\partial_{\bar{\phi}}\bar{W} 
- \frac{\alpha \bar{W}(1 -i \kappa \partial_{\bar{\phi}}f)}{\phi - \bar{\phi}-i\kappa f} \right] - 3 \frac{\bar{W}W}{M_{\rm Pl}^2}\right\rbrace 
\eea

\subsection{Taylor and Laurent series expansions}

We now proceed to analyze the perturbative expansion of superpotentials, based on the K\"{a}hler potentials analyzed in the previous subsections. 
Our goal is to consider inflation in these models, so we will use the known experimental constraints on inflation. 

In order to do that, we must calculate the standard slow-roll parameters $\epsilon$ and $\eta$, 
\bea
\epsilon&=&\frac{M_{\rm Pl}^2}{2}\left(\frac{1}{V}\frac{dV(\phi_{\rm can})}{d\phi_{\rm can}}\right)^2\cr
\eta&=&M_{\rm Pl}^2\left(\frac{1}{V}\frac{d^2 V(\phi_{\rm can})}{d\phi_{\rm can}^2}\right)\;,
\eea
where $\phi_{\rm can}$ is the {\em canonical scalar } (inflaton), and one should compare the resulting scalar tilt $n_s$ and tensor to scalar ratio $r$, given by 
\be
n_s-1=-6\epsilon+2\eta;\;\;\;\;
r=16\epsilon\;,
\ee
with the experimental results, the $n_s$ from the 2015 Planck paper \cite{Ade:2015lrj}, 
and the bound on $r$ from the 2015 joint Planck-BICEP2 paper \cite{Ade:2015tva}, 
\be
n_s-1=-0.032\pm 0.006;\;\;\;\; r<0.12.
\ee
In the case that $\eta\ll \epsilon$, it means that we need
\be
\epsilon\sim 5\times 10^{-3}.
\ee

Note that for a real K\"{a}hler metric $g_{\phi\bar\phi}=\d_\phi\d_{\bar\phi}K$ and imaginary inflaton, since
\be
\frac{1}{2}(d\phi_{\rm can})^2=g_{\phi\bar\phi} |d\phi|^2=g_{\phi\bar\phi}d\sigma^2\;,
\ee
we have
\bea
\epsilon&=&\frac{M_{\rm Pl}^2}{4 g_{\phi\bar\phi}}\left(\frac{1}{V}\frac{dV}{d\sigma}\right)^2\cr
\eta&=&\frac{M_{\rm Pl}^2}{2}\left(\frac{1}{V}\frac{d}{\sqrt{g_{\phi\bar\phi}}d\sigma}\left(\frac{d}{\sqrt{g_{\phi\bar\phi}}d\sigma}V\right)\right).
\eea

{\bf The eta problem for small field inflation}

Note that since we need $\epsilon\ll 1$ and $\eta\ll 1$, if in small field inflation the potential is Taylor expanded near the origin as
\be
V\simeq A\left(1+C_1\frac{\phi_{\rm can}}{M_{\rm Pl}}+C_2\frac{\phi_{\rm can}^2}{M^2_{\rm Pl}}+...\right)\;,
\ee
then we need to have $C_1\ll 1, C_2\ll 1$. But generically we would get $C_1,C_2\sim 1$, which would not give inflation. This is the eta problem 
for small field inflation. To obtain inflation, we will need to either {\em impose} $C_1,C_2\ll 1$, or simply for them to vanish, and only have 
higher orders in $\phi_{\rm can}$. 

{\bf Other constraints}

One should also fix the normalization of the potential using the WMAP data \cite{wmap9}
\be
\frac{H^2_{\rm infl}}{8\pi^2\epsilon M_{\rm Pl}^2}\simeq 2.4\times 10^{-9}\;,
\ee
where $H^2_{\rm infl}=V_{\rm infl}/(3M_{\rm Pl}^2)$, as well as the number of e-folds, 
\be
N_e=\int_{\phi_{\rm can,initial}}^{\phi_{\rm can, final}}\frac{d\phi_{\rm can}/M_{\rm Pl}}{\sqrt{2\epsilon}}=\int_{\sigma_i}^{\sigma_f}
\frac{d\sigma\sqrt{2g_{\phi\bar\phi}}/M_{\rm Pl}}{\sqrt{2\epsilon}}.
\ee

Note that if we are interested only in the {\em leading} behaviour in $\sigma$ for $\epsilon$ and $N_e$, we can neglect the subleading terms 
in $g_{\phi\bar\phi}$ for the calculation of $\epsilon, \eta, N_e$, 
since $g_{\phi\bar\phi}$ appears outside derivative terms, so they would only contribute to subleading terms in the quantity calculated.
In the following, we again revert to $M_{\rm Pl}=1$.

\subsubsection{Perturbative K\"{a}hler potential}

We start by considering the case of K\"{a}hler potential that is canonical plus corrections. As we saw, corrections up to fourth 
order in $\Phi+\bar\Phi$ are equivalent to corrections to second order in $\bar\Phi\Phi$, so we consider
\be
K(\Phi, \bar{\Phi}) \simeq \Phi\bar{\Phi} + \kappa(\Phi\bar{\Phi})^2\;,
\ee
understood as an approximation in $\kappa$. 
Then for a general superpotential $W$, the scalar potential becomes 
\bea
V &\simeq& e^{\phi\bar{\phi}+\kappa(\bar\phi\phi)^2}
\left[(1-4\kappa|\phi|^2)\left\{\left|\partial_{\phi}W +W\bar{\phi} \right|^2 +2\kappa\bar\phi\phi\left(\bar{W}\phi\partial_{\phi}W 
+ \right.\right.\right.\nonumber\\
&&\left.\left.\left.+W\bar{\phi}\partial_{\bar{\phi}}\bar{W} + 2W\bar{W}\bar{\phi}\phi + 2\kappa\phi \bar{\phi}|\phi|^2|W|^2 \right)\right\} -3|W|^2 
+{\cal O}(\kappa^2)\right]\cr
&\simeq& e^{\bar\phi\phi}\left\{\left|\d_\phi W+\bar\phi W\right|^2-3|W|^2+\kappa|\phi|^2\left[2\phi\bar W\d_\phi W +2W\bar\phi \d_{\bar\phi}\bar W
+4|\phi|^2|W|^2\right.\right.\cr
&&\left.\left.-4\left(1-\frac{\bar\phi\phi}{4}\right)\left|\d_\phi W+\bar\phi W\right|^2-3\bar\phi\phi|W|^2\right]+{\cal O}(\kappa^2)\right\}.
\eea

{\bf Laurent expansion of superpotential}

Consider the case that the superpotential starts with a pole (negative power), and is Laurent expandable, or rather, a pole times a Taylor expandable
function. We will consider the Taylor expansion up to fourth order, i.e. 
\be
W(\Phi) = \Phi^{-n}(a + b\Phi + c\Phi^2 +d\Phi^3 +f\Phi^4).\label{wminusn}
\ee
On $\phi=i\sigma$, {\em if $a,b,c,d,f$ are all real}, we obtain
\be
V\simeq a^2e^{\sigma^2}\left(n^2 \sigma^{-2n -2} + A \sigma^{-2n} + B\sigma^{-2n + 2} \right),  
\ee
where we have defined
\bea
a^2A &=& b^2+ (n^2-2n)(b^2-2ac)-a^2(2n+3+4\kappa n^2)\nonumber\\
a^2B &=& a^2(1 +4\kappa n+\kappa n^2) + (4c^2-6bd)+ (n^2-4n)(c^2 + 2af - 2bd)\nonumber\\
&&-(2n +1 + 4\kappa n^2 -8\kappa n)(b^2 - 2ac)-4\kappa b^2.
\eea

Note that we have considered also the second subleading term with coefficient $B$, but its expression is correct only if $n=0$, so that the 
first term vanishes. Otherwise, since $K$ is correct only up to the first nontrivial order, the final result for $V$ should also be correct only for the 
first subleading term. 
We next calculate the slow-roll parameter $\epsilon$. As we said, for its calculation we need to keep only the leading term in $g_{\phi\bar\phi}$,
so we approximate
\be
g_{\phi\bar\phi}\simeq 1+4\sigma^2\simeq 1\;,
\ee
so that $\phi_{\rm can}\simeq \sqrt{2}\sigma$. Then we obtain
\be
\epsilon = \frac{1}{2}\left(\frac{1}{V}\frac{d\sigma}{d\phi_{\rm can}}\frac{dV}{d\sigma}\right)^2 
\simeq\frac{1}{4}\left\lbrace2\sigma + \frac{1}{\sigma}\frac{-n^2(2n+2) - 2nA\sigma^2 -(2n-2)B\sigma^4}{n^2+ A\sigma^2+B\sigma^4}\right\rbrace^2.
\ee

If $n\neq 0$ and $n\neq -1$, then we have		
\begin{equation}
\epsilon \simeq \frac{1}{4\sigma^2}\left(2n+2\right)^2 \gg 1,
\end{equation}	
which contradicts our assumptions, since we want $\epsilon\ll 1$, so this case does not give inflation.

If $n=0$, we get		
\begin{equation}
\epsilon \simeq \frac{1}{4}\left(2\sigma + \frac{2B\sigma^4}{A\sigma^3}\right)^2 = \sigma^2\left(1 + \frac{B}{A}\right)^2,
\end{equation}
where now $a^2A = b^2-3a^2$ and $a^2B = a^2 -(b^2-2ac) + 4c^2-6bd-4b^2\kappa$. Now we can adjust the parameters of the superpotential in order 
to get $\epsilon \ll 1$. Next we need to calculate the number of e-folds $N_e$ and put the experimental constraints of having at least 60 
e-folds of inflation. Assuming that we can set $\frac{B}{A} \sim 1$, we have
\begin{equation}
N_e = \int^{\phi_{\rm can, f}}_{\phi_{\rm can,in}}\frac{1}{\sqrt{2\epsilon}}d\phi_{\rm can}/M_{\rm Pl} 
=\int^{\sigma_f}_{\sigma_i}\frac{1}{\sqrt{2}\sigma\left(1+\frac{B}{A}\right)}
d\sigma \simeq \int^{\sigma_f}_{\sigma_i}\frac{1}{2\sqrt{2}\sigma}d\sigma.
\end{equation}
The constraint of $N_e\geq 60$ implies $\frac{\sigma_f}{\sigma_i} \geq e^{120\sqrt{2}}\simeq 5\times 10^{73}$, 
which is an incredible fine-tuning: it means that $\sigma_i$ needs to 
be incredibly close to zero for inflation to happen.  

If $n=-1$, we have
\begin{equation}
\epsilon \simeq \frac{1}{4}\left(2\sigma + \frac{1}{\sigma}2A\sigma^2\right)^2 = \sigma^2\left(1 + A\right)^2,
\end{equation}
and this leads to the same fine tuning for generic $A$. Note that $n=-1$ means the leading term in $W$ is linear, $W=a\phi$, a term that is known 
to give inflation. 

However, if we have $A=-1$, then the leading term vanishes, so we must consider the next power in $\epsilon$, obtaining
\begin{equation}
\epsilon \simeq \frac{1}{4}\left[(4B -2 )\sigma^3 \right]^2. 
\end{equation}
Then the potential becomes approximately
\be
V\simeq a^2[1+(B-1/2)\sigma^4+{\cal O}(\sigma^6)]\;,
\ee
so we also have for the other slow-roll parameter
\be
\eta\simeq 3(2B-1)\sigma^2\gg \epsilon\;,
\ee
but it is still $\ll 1$, so we satisfy the slow-roll conditions. 

Now we have from the experimental value of $n_s-1$
\be
n_s-1\simeq 2\eta \sim -3\times 10^{-2}\;,
\ee
which implies 
\be
3(2B-1)\sigma_f^2\sim -3/2\times 10^{-2}\;,
\ee
and it constrains 
\be
\sigma_f\sim \frac{1}{\sqrt{2-4B}}\times 10^{-1}.
\ee

This would lead to a reasonable $\sigma_f/\sigma_i$, since now
\be
N_e=\frac{1}{2(2B-1)}\left(\frac{1}{\sigma_i^2}-\frac{1}{\sigma_f^2}\right)\simeq \frac{1}{2(2B-1)}\frac{1}{\sigma_i^2}\leq 60\;,
\ee
which constrains $\sigma_i$. 

So we need to find the solutions to the $A=-1$ constraint, since they are the cases that give good inflation.

For $n=-1$, we have 
\be
a^2A=4b^2-a^2-6ac -4a^2\kappa=-a^2\Rightarrow 4b^2/a^2-6c/a-4\kappa^2=0.
\ee
This is one equation constraining 3 parameters $b,c,\kappa$, with $d,f$ unconstrained (but they correspond to higher order coefficients 
in the $\Phi$ expansion of the superpotential). 

We see that a simple solution is $b=0, c=0,\kappa=0$, though it isn't the most general. In that case
\be
a^2B(b=c=\kappa=0,n=-1)=a^2+10af\;,
\ee
and we need to impose that $B\neq 1/2$, i.e. $f/a\neq -1/20$. 

Such a model will be considered in more detail below. In fact, it
has already  been considered by Izawa and Yanagida \cite{Izawa:1996dv}. 

From the previous analysis, we need to generalize to the case with general complex coefficients (instead of real ones), 
but for ease of calculations we consider the case with a nontrivial subleading term with a general power, i.e.
\be
W(\Phi)=\Phi^{-n}(a+b\Phi^m)\;,
\ee
where $a,b$ are complex. Then we obtain 
\be
D_\phi W\simeq \frac{a}{\Phi^{n+1}}(-n+\bar \Phi \Phi +2\kappa (\bar\Phi \Phi)^2)+\frac{b\Phi^m}{\Phi^{n+1}}(m-n+\bar \Phi \Phi+2\kappa (\bar \Phi\Phi)^2)\;.,
\ee
and the potential is 
\bea
V(\Phi=i\sigma)&\simeq& \frac{1+\sigma^2}{\sigma^{2n+2}}\left[\frac{({\rm Re}\; a(-n+\sigma^2+2\kappa \sigma^4)+{\rm Re}(b i^m)\sigma^m(m-n+\sigma^2
+2\kappa \sigma^4))^2}{1+4\kappa\sigma^2}\right.\cr
&&\left.+\frac{+({\rm Im}\; a(-n+\sigma^2+2\kappa \sigma^4)+{\rm Im}(b i^m)\sigma^m(m-n+\sigma^2
+4\kappa \sigma^4))^2}{1+4\kappa\sigma^2}\right.\cr
&&\left.-3\sigma^2\left[({\rm Re}\; a+{\rm Re}\; (bi^m)\sigma^n)^2+({\rm Im}\; a+{\rm Im}\; (bi^m)\sigma^n)^2\right]\right].
\eea

We see that nothing new is obtained in this more general case, so the above analysis was complete. 

{\bf Taylor expansion of superpotential}

One can consider next the case that the superpotential is Taylor expandable, i.e. of the form (\ref{wminusn}), but with $n<0$. 
We realize that the analysis did not depend on $n$ being positive or negative, in fact the only case we have found that has inflation had $n=-1$,
so the same conclusion applies: the model with $n<0$ has
\begin{equation}
\epsilon \simeq \frac{1}{4\sigma^2}\left(2n+2\right)^2 \gg 1,
\end{equation}
except in the case $n=-1$, which means that we do not get slow-roll inflation unless we have a leading linear potential. 

Even in the case of the linear superpotential however, we need to be concerned about the {\em sign} of the first correction away from 
the plateau: i.e., if the potential becomes of the type $V=A(1+C_1\sigma^n+...)$ with $C_1>0$, $n>0$ then while we have inflation, we might not be able to 
{\em end it}, since this will lead us towards the local minimum at $\sigma=0$, which has nonzero value for the potential. This is the case if 
we choose simply $K=\bar \Phi \Phi$ and $W=\Phi$, 
when we have
\be
V\simeq 1+\sigma^4+...
\ee
as we can easily check with the above formulae. But if instead we have $V=A(1-C_1\sigma+...)$
with $C_1>0$, then we are led {\em away} from the local maximum of the potential at $\sigma=0$.

We have therefore analyzed all possibilities for the case of a potential that is Taylor expandable except for poles. 

Note that we could consider also combinations of exponentials and polynomials, but these cases are also Taylor expandable. 
For instance, 
\bea
W_1&\equiv&(1+a\Phi^n)e^{b\Phi^m}\simeq 1+a\Phi^n+b\Phi^m+\frac{b^2}{2}\Phi^{2m}+ab\Phi^{n+m}+...\cr
W_2&\equiv& e^{a\Phi^n+b\Phi^m}\simeq 1+a\Phi^n+b\Phi^m+\frac{a^2}{2}\Phi^{2n}+\frac{b^2}{2}\Phi^{2m}+ab\Phi^{n+m}+...\;,
\eea
and the only issue is how many terms should we keep in the expansion for the consistency of the approximation. For instance, for $n=1$ and $m=2$, 
we should keep a subleading $\Phi^2$ correction in $dW/d\Phi$, that is, we should keep including terms of $\Phi^{3n}=\Phi^3$ and $\Phi^{n+m}=\Phi^3$, 
but not $\Phi^{2m}=\Phi^4$ terms. 

But exponentials are useful tools for organizing these Taylor expansions. For instance, considering the superpotential (see \cite{Hinterbichler:2013we})
\be
W(\Phi)=\Phi e^{ib\Phi}\;,
\ee
with $b$ real and positive, we find the potential
\be
V\simeq 1-4b\sigma+7b^2\sigma^2\;,
\ee
which does move perturbatively away from $\sigma=0$, though one would have to check nonperturbatively if it drops all the way to zero for large enough 
$\sigma$. For this simple model we find
\be
\epsilon=4b^2;\;\;\;\; \eta=\frac{7}{2}b^2\;,
\ee
so it seems like we have a good small field inflation if $b\ll1$. However, we have to consider also the number of e-folds, which becomes
\be
N_e=\int_{\sigma_i}^{\sigma_f}\frac{d\sigma}{\sqrt{\epsilon}}=\frac{\sigma_f-\sigma_i}{2b}\;,
\ee
and has to be larger than about 50-60. We can achieve all these constraints with small enough $b$, however if we also want to impose the 
experimental result for $n_s-1\simeq 0.032$, with $n_s-1=-6\epsilon+2\eta$, we obtain
\be
-17 b^2\simeq -0.032\Rightarrow b\simeq \frac{1}{22.5}\;,
\ee
and then the number of e-folds constraint implies $\sigma_f-\sigma_i\gsim 5$, which contradicts the constraint to be in the small inflaton region. 

But the next simplest case of exponential solves this problem. Consider now
\be
W=\Phi e^{b\Phi^2}\simeq \Phi(1+b\Phi^2+{\cal O}(\Phi^4))\;,
\ee
with $b$ real and positive. Then we find the potential
\be
V\simeq 1-6b\sigma^2\;,
\ee
which has the required decreasing plateau, and gives 
\be
\epsilon\simeq b^2\sigma^2;\;\;\;\;
\eta=-2b\;,
\ee
so, since $|\eta|\gg \epsilon$, we have $n_s-1\simeq 2\eta\simeq -2b$, and matching against the experimental result of $-0.032$, we get
\be
b\simeq \frac{1}{60}.
\ee
In turn, that means we can easily fix the number of e-folds, since now
\be
N_e=\int_{\sigma_i}^{\sigma_f}\frac{d\sigma}{\sqrt{\epsilon}}=\frac{1}{2b}\ln\frac{\sigma_f}{\sigma_i}\simeq 120\ln \frac{\sigma_f}{\sigma_i}\;,
\ee
which can easily be made to be larger than 60.

The remaining possibility is of an essential singularity in the superpotential at $\Phi=0$, or a nonperturbative contribution.

{\bf Logarithmic superpotential}

The first simple possibility of an essential singularity (which does not reduce to a pole or finite sum of poles) at $\Phi=0$ is a log superpotential, i.e.
\be
W(\Phi)=A\ln\Phi.
\ee
Then the potential is 
\bea
V(\phi) &\simeq & A^2e^{\phi\bar{\phi}} \left[\frac{\left(\frac{1}{\phi}+(1+2\kappa \phi\bar{\phi})\bar{\phi}\ln\phi\right)\left(\frac{1}{\bar{\phi}}
+(1+2\kappa\phi\bar{\phi})\phi\overline{\ln\phi}\right)}{(1+4\kappa\phi\bar{\phi})} - \right. \nonumber\\
&&\left. -3\left| \ln\phi\right|^{2} \right]\cr
&\simeq &A^2e^{\sigma^2}\left(\frac{1}{\sigma^2}-3\ln^2\sigma\right).
\eea
Note that for consistency we should only keep terms up to the first nontrivial subleading term in $\sigma^2$. The slow-roll parameter $\epsilon$ is 
then found to be
\be
\epsilon\simeq \frac{1}{\sigma^2}\gg 1\;,
\ee 
so we don't get inflation.

{\bf Exponential superpotentials with essential singularity}

Consider next a superpotential of the general form
\be
W=Ae^{a\Phi^{-n}(1+b\Phi^m)}\;,
\ee
and consider Re$(a\; i^{-n})>0$. The the superpotential at $\Phi=i\sigma$ has an essential singularity as $\sigma\rightarrow 0+$, 
as $e^{{\rm Re}(a i^{-n})\sigma^{-n}}\rightarrow\infty$. 

We obtain 
\be
D_\Phi W=Ae^{a\Phi^{-n}(1+b\Phi^m)}\left(-a\Phi^{-n-1}(n+b\Phi^m(n-m))+\bar\Phi(1+2\kappa\bar \Phi\Phi)\right)\;,
\ee
and for $\Phi=i\sigma$ we obtain the potential
\be
V\simeq A^2 e^{F \sigma^{-n}+G \sigma^{m-n}}|a|^2 \sigma^{-2n-2}(n^2+n(n-m)\sigma^m H+(m-n)^2|b|^2\sigma^{2m}+(1-4\kappa)n^2\sigma^2)\;,
\ee
where we have defined the quantities
\be
F\equiv ai^{-n}+a^*(-i)^{-n};\;\;\;\;
G\equiv ab i^{m-n}+a^*b^*(-i)^{m-n};\;\;\;
H\equiv b i^m+b^*(-i)^m.
\ee
Note that per our assumption, we have $F>0$. Then the potential is rapidly varying near $\sigma=0$, the opposite of flat that we need for 
inflation, so we don't get inflation. Specifically, we have
\be
V\sim e^{F\sigma^{-n}}\sigma^{-2n-2}\Rightarrow \epsilon=\frac{M_{\rm Pl}^2}{2}\left(\frac{V'}{V}\right)^2\sim \frac{1}{2}\left(\frac{-nF}{\sigma^{n+1}}-\frac{2n
+2}{\sigma}+...\right)^2\gg 1.
\ee

{\bf Nonperturbative contribution}

If on the other hand, one chooses Re$(ai^{-n})<0$, the superpotential at $\Phi=i\sigma$ is nonperturbative: it obeys $W(\sigma=0)=0$, 
but it is not Taylor expandable around $\sigma=0$. 

But the formulas above are still valid, just that now we have $F<0$. We still obtain $\epsilon\gg 1$, so no inflation.

{\bf Special inflationary models}

In this way, we have analyzed simple cases of singularities of $W$ at $\Phi=0$. Of course, we can always consider more complicated singularites, 
that mix exponentials, logs and powers in the leading term. For instance, one could consider the combination
\be
W=\exp[\b e^{ia/\Phi}]\;,
\ee
with $a,b\in \mathbb{R}_+$, that has an essential singularity at $\Phi=0$, but of a more complicated type than the one above. 

Or, one could consider more complicated nonperturbative contributions, like the case considered in \cite{Hinterbichler:2013we}, of 
\be
W=\exp[-\b e^{-ia/\Phi}]\;,
\ee
with $a,b\in\mathbb{R}_+$, that becomes equal to 1 at $\Phi\rightarrow i0$.

But the aim here was to classify all natural possibilities, leaving a choice of highly special ones like the above to always be a possibility.

\subsubsection{Logarithmic K\"{a}hler potential}

Consider next the logarithmic K\"{a}hler potentials in (\ref{logk}). More concretely, we will consider corrections both multiplying the log and 
multiplying the factor inside the log, i.e.,
\be
K=-\a(1+\gamma\bar\Phi\Phi)\ln [-i(\Phi-\bar\Phi)(1+\b\bar\Phi\Phi)].
\ee
Then 
\be
\d_\Phi K\simeq \frac{\a}{\bar\Phi-\Phi}(1+\gamma\bar\Phi \Phi)-\a\gamma\bar\Phi \ln[-i(\Phi-\bar\Phi)]-\a\b\bar\Phi\;,
\ee
and 
\be
g_{\Phi\bar\Phi}(\Phi=i\sigma)\simeq \frac{\a}{4\sigma^2}[1-(4\b+3\gamma)\sigma^2-4\gamma\sigma^2\ln(2\sigma)].
\ee
We see that we need $\a>0$ in order not to have a ghost ($g_{\Phi\bar\Phi}>0$).

{\bf Taylor and Laurent expansion of the superpotential}

Consider the generic superpotential
\be
W=A\Phi^n(1+b\Phi^m)\;,
\ee
where $m>0$, but $n$ can be either positive or negative, obtaining either a Taylor expansion, or a Taylor expansion around a pole (Laurent expansion). 
Then  we obtain the potential 
\bea
V&\simeq& \frac{4A^2\sigma^{2n-\a}}{\a 2^\a}(1-\a\b \sigma^2-\a\gamma\sigma^2\ln(2\sigma))
\left\{\left[\left(n-\frac{\a}{2}\right)^2-\frac{3\a}{4}\right]+\sigma^2\left[\left(n-\frac{\a}{2}\right)^2(4\b+3\gamma)
\right.\right.\cr
&&\left.\left.-\a\left(n-\frac{\a}{2}\right)(\gamma+2\b)+\gamma\ln(2\sigma)((2n-\a)^2-\a(2n-\a))\right]\right.\cr
&&\left.
+\sigma^mF\left[\left(n-\frac{\a}{2}\right)\left(m+n-\frac{\a}{2}\right)-\frac{3\a}{4}\right]+\sigma^{2m}|b|^2\left[\left(m+n-\frac{\a}{2}\right)^2-\frac{3\a}{4}\right]
\right\}\;,\cr
&&
\eea
where $F=bi^m+b^*i^{-m}$. Note that if $m=1$ we need to keep all these terms, if $m=2$ we can drop the $\sigma^{2m}$ term, and if $m>2$ 
we can drop $\sigma^m$ and $\sigma^{2m}$. We see that we need $\a=2n$ in order to have inflation, since a power law will not give inflation. 
But then we see that the plateau is actually an AdS plateau (at negative potential), so it will again not give inflation. 

Therefore we cannot obtain inflation in this scenario. We must again move to superpotentials with essential singularities, and nonperturbative ones. 

{\bf Logarithmic superpotential}. 

Consider the superpotential
\be
W=A\ln(-i\Phi).
\ee
Then we obtain the potential
\be
V\simeq \frac{4A^2\sigma^{-\a}}{\a 2^\a}(1-\a\b\sigma^2-\a\gamma\sigma^2\ln(2\sigma))\left[\left(1-\frac{\a}{2}\ln\sigma\right)^2-\frac{3\a}{4}\ln^2\sigma
+{\cal O}(\sigma^2\ln\sigma)\right].
\ee
We see that we can only obtain a plateau if $\a=0$ (a power law potential has no plateau), but this is forbidden, since then we have no K\"{a}hler 
potential. Therefore also in this case we cannot obtain new inflation. 

{\bf Exponential superpotentials with essential singularities}

Consider now a superpotential
\be
W=Ae^{a\Phi^{-n}(1+b\Phi^m)}\;,
\ee
with $n>0$ (since otherwise we can make a Taylor expansion). Then 
\bea
D_\Phi W&=&Ae^{a\Phi^{-n}(1+b\Phi^m)}\left[-a\Phi^{-n-1}(n+b(n-m)\Phi^m)\right.\cr
&&\left.+\frac{\a}{\bar\Phi-\Phi}(1+\gamma\bar\Phi\Phi)-\a\b\bar\Phi-\a\gamma
\bar\Phi\ln[-i(\bar\Phi-\Phi)]\right]\;,
\eea
and the resulting potential is 
\be
V\simeq \frac{4A^2e^{F\sigma^{-n}+G\sigma^{m-n}}}{\a 2^\a \sigma^{\a+2n}}(1-\a\b\sigma^2-\a\gamma\sigma^2\ln(2\sigma))\left[|a|^2n^2-\frac{3\a}{4}
\sigma^{2n}+{\cal O}(\sigma^2\ln\sigma)+{\cal O}(\sigma^m\ln\sigma)\right].
\ee
Since we assumed $n>0,\a>0$, the potential is dominated by the 
\be
\frac{e^{F\sigma^{-n}+G\sigma^{m-n}}}{\sigma^{\a+2n}}
\ee
factor, which we saw when analyzing the canonical K\"{a}hler potential  that does not give inflation either. And again the result is independent on 
the sign of $F$, so it applies both in the case of a singularity at $\sigma=0$ and of a nonperturbative contribution. 

In conclusion, we cannot obtain inflation with the logarithmic K\"{a}hler potential either.

\subsubsection{Linear term plus corrections and a general supergravity embedding}

Finally, we consider a case for the K\"{a}hler potential that would be a bit counterintuitive on first thought, yet it contains a very important
model, that allows us to embed a general inflationary model inside minimal supergravity. 

We considered the perturbative (Taylor expansion) K\"{a}hler potential as starting with the canonical one, $K=\bar\Phi\Phi+...$, but that 
is not necessary, as we said at the beginning of the section. We could have a $K$ that starts with the linear term $\a(\bar \Phi+\Phi)$, and 
continues with quadratic and higher corrections. It is not clear how such a $K$ would be obtained in perturbation theory, but we need to 
consider it, since it contains  a very important example, considered in \cite{Ketov:2014qha,Ketov:2014hya}. 

Consider the exact K\"{a}hler potential 
\bea
K&=&-3\ln\left(1+\frac{\bar\Phi+\Phi}{\sqrt{3}}\right)\cr
&\simeq & -\sqrt{3}(\bar \Phi+\Phi)-\frac{1}{2}(\bar\Phi+\Phi)^2+...\;,\label{specialkahler}
\eea
where in the second line we have put the expansion only to the first subleading order. Note that to this order, we could make a K\"{a}hler 
transformation and get rid of the $f(\Phi)+g(\bar \Phi)$ terms, and remain with $K=-\bar\Phi \Phi+...$, but it is important that we have higher 
orders. 

The canonical inflaton is $\phi_{\rm can}=\sqrt{2}{\rm Im}\Phi$, and the real part is stabilized at zero. Then it is easy to see that the 
kinetic term of $\phi_{\rm can}$ is in fact canonical, and the potential becomes
\be
V(\phi_{\rm can})=|\d_\Phi W(i{\rm Im}\Phi)|^2=(\hat W'(\phi_{\rm can}))^2\;,
\ee
where 
\be
W(\Phi)=\frac{1}{\sqrt{2}}\hat W(-\sqrt{2}i\Phi)
\ee
and $\hat W(x)$ is a {\em real} function. 
Reversely, that means that for any positive definite potential, so that it can be written as a total square, one can find a superpotential that reproduces 
it. Thus most inflationary models can be embedded in minimal supergravity in this way.

\subsection{Constraints from larger field}

In order to really have inflation, we should be able to {\em end it}, which means that we should impose that the potential not only starts at 
a plateau, and starts to go down it, but that it is not followed (nonperturbatively) by a local maximum, but rather by a minimum. It could happen 
that after an initial downhill period, the potential grows to a local maximum, and then settles to a minimum. This would be the case in the "old 
inflation" type models, which however we know that don't work in detail, since one needs to tunnel through the maximum. 

We have seen that a logarithmic K\"{a}hler potential is ruled out (does not give inflation) at the first stage, and we will ignore the 
special case (\ref{specialkahler}), which can embed any positive definite potential in supergravity, since its K\"{a}hler potential is 
very special. Then the only possibility was a K\"{a}hler potential that is a Taylor expansion around the canonical one, and moreover 
in that case we only obtained inflation for the special Taylor-expandable superpotential that starts off with a linear term.
We now must consider how to continue this model to larger fields, and find a potential where the plateau is followed by a minimum.

One example of such a model, that does satisfy these constraints, is the model considered by Yzawa and Yanagida \cite{Izawa:1996dv}
(see also the review \cite{Yamaguchi:2011kg}), with
\bea
K&=&\bar \Phi\Phi\cr
W&=&v^2\Phi-\frac{g}{n+1}\Phi^{n+1}\label{yy}
\eea
where $n\geq 3$, and also $v\ll 1, g\sim 1$. The potential becomes
\be
V(\Phi=i\sigma)=e^{\sigma^2}\left\{\left|v^2(1+\sigma^2)-gi^n\sigma^n\left(1+\frac{\sigma^2}{n+1}\right)\right|^2-3\sigma^2\left|v^2
-\frac{g}{n+1}i^n\sigma^n\right|^2\right\}
\ee

Because of the condition $g/v^2\gg 1$, we can approximate the potential, down to its minimum, and for a short while after it as well, by 
writing $K\simeq 0$, $D_\Phi W\simeq v^2-g\Phi^n$, $W\simeq 0$, i.e.
\be
V\simeq v^4-2v^2g\sigma^n-g^2\sigma^{2n}\;,
\ee
which has a minimum at 
\be
\sigma_m\simeq \left(\frac{v^2}{g}\right)^{\frac{1}{n}}.
\ee
Note that the negative value of the potential  at the minimum is obtained by allowing back in some of the neglected terms, namely 
\be
V_{min}\simeq -3e^{\sigma^2}|W(\sigma_{min})|^2.
\ee

However, as can be easily seen, this potential grows without bound after the negative (AdS) minimum. But if we are not interested in reaching 
large field values, we are fine. Otherwise, we need to extend it to higher values of the fields by adding extra terms. 

If the minimum is not required to be close to zero, we can consider even the model of section 2.2, with $K=\bar\Phi\Phi$ and 
$W=\Phi e^{b\Phi^2}$, whose exact potential is 
\be
V=e^{(1-2b)\sigma^2}[1-(1+4b)\sigma^2+\sigma^4(1-2b)^2].
\ee
From the experimental inflationary constraints, we saw that we needed $b\simeq 1/60$, and we can neglect it in the above potential.
Then the square bracket has a minimum at $x=\sigma^2=1/2$, of value $3/4$, compared with 1 at $\sigma=0$, and the true minimum of the 
potential, including the exponential, is nearby.  This potential then does have a local 
minimum, however at a rather large and positive value of the potential, and then it goes to infinity asymptotically. It is not good for 
phenomenology, since it will give a too large cosmological constant, since the field would be stuck at its (positive energy) minimum. 

In conclusion, potentials satisfying our constraints are possible, and moreover the constraints themselves are nontrivial, and restrict to a very 
small set of models, of canonical K\"{a}hler potential (maybe plus corrections), and a linear superpotential plus corrections of higher order.

\section{Constraints from large field and small potential}

Finally, we consider the constraint that the potential needs to remain small after the minimum, and moreover go down to zero asymptotically. 
As we mentioned, in the Izawa and Yanagida model of last subsection, this was not the case, and the potential grows without bounds after the negative 
(AdS) minimum. Yet at least perturbatively that was the only possibility we have found, so we need to analyze possible nonlinear completions for it that 
can give the desired result. 

Consider first the possibilities for the K\"{a}hler potential. We want to obtain a potential that goes to zero asymptotically, but the ${\cal N}=1$
supergravity formula has an overall $e^K$ factor, whereas in the rest we have only $g_{\bar\Phi\Phi}$, $\d_\Phi K$ and the superpotential 
and its derivatives. That means that $K$ being nonlinearly a growing exponential  $e^{a(\bar\Phi\Phi)^\b}$ is ruled out: we would get an 
$e^{e^{a\sigma^{2\b}}}$ growth, that could only be compensated by a superpotential that decays even faster, $e^{-e^{a\Phi^{2\b}}}$, which 
seems very unlikely from a physics point of view, and also can be explicitly shown to not give inflation in any case. 

One could consider $K$ to be instead a decaying exponential, like for instance
\be
K=\a\bar\Phi\Phi e^{-\b\bar\Phi\Phi}\;,
\ee
but in this case we obtain a K\"{a}hler metric (giving the kinetic term fot the inflaton)
\be
g_{\bar\Phi\Phi}=\d_\Phi\d_{\bar\Phi}K=\a e^{-\b\bar\Phi\Phi}[1-3\b\bar\Phi\Phi+\b^2(\bar\Phi\Phi)^2]\;,
\ee
which becomes negative for a region. Indeed, we can check that, considering the variable $x=\b\bar\Phi\Phi$, the square bracket has a 
minimum at $x=3/2$, where it takes the value $-5/4$. But a negative metric means the inflaton becomes a ghost, which is forbidden. 

We encounter similar problems with nonlinear completions of the K\"{a}hler potential that go to zero too fast (such as to compensate for 
the asymptotic growth due to the superpotential). 

A polynomial K\"{a}hler potential $P_n(\Phi)$ at infinity 
means that $e^K$ becomes $e^{P_n(\Phi)}$, which would need to be compensated by a superpotential that is exponentially decaying at infinity.
The example from section 2.2, with $K=\bar\Phi\Phi$ and $W=\Phi e^{b\Phi^2}$ was of this type. In fact this model to does reach a local 
minimum as we saw in the last section, but we need it also to go down asymptotically at infinity, so it needs to be further corrected. 
Similarly, the better defined Izawa-Yanagida model has a local AdS minimum close to zero, but then it grows without bounds, so 
it must be modified at large values of the field. 

\subsection{Trial modification of the Izawa-Yanagida model}

Consider the following modification of the Izawa-Yanagida model with $n=4$, 
\bea
K&=&\a\Phi\bar\Phi\cr
W&=&e^{i^p\b \Phi^m}\left(\Phi-\frac{\tilde g}{5}\Phi^5\right)\;,
\eea
where $\tilde g$ stands for what we called $g/v^2$ in the unmodified model. This $\tilde g$  is taken to be $\gg 1$, and we also choose
$i^{p+m}=-1$. Then the potential is
\be
V=\frac{e^{\a\sigma^2-2\b \sigma^m}}{\a}\left\{\left[1-\tilde g\sigma^4+\left(1-\frac{\tilde g}{5}\sigma^4\right)(-m\b\sigma^m+\a\sigma^2)\right]^2
-3\a\sigma^2\left(1-\frac{\tilde g}{5}\sigma^4\right)^2\right\}.
\ee
Consider moreover $m=4$, $p=2$ and $\b\gg 1$. Then 
\be
V=\frac{e^{\a\sigma^2-2\b\sigma^4}}{\a}\left\{\left[1-\tilde g\sigma^4+\left(1-\frac{\tilde g}{5}\sigma^4\right)(-4\b \sigma^4+\a\sigma^2)\right]^2
-3\a\sigma^2\left(1-\frac{\tilde g}{5}\sigma^4\right)^2\right\}\;,
\ee 
and then before the minimum, and a bit after, we can approximate the potential as 
\be
V\simeq \frac{e^{\a\sigma^2-2\b \sigma^4}}{\a}\left\{1-\a\sigma^2
-2(\tilde g+4\b)\sigma^4+{\cal O}(\sigma^6)\right\}.
\label{potapprox}
\ee
Note that the $\sigma^2$ terms cancel in the expansion (after expanding the exponential), so the first term is at order $\sigma^4$, so we would 
get good inflation, and moreover the term starts to drop towards a minimum. Also $V(\sigma\rightarrow \infty)=0$, but unfortunately, 
a scan of parameter space shows that we cannot find a situation with a local minimum close to zero, and no maximum higher than 
the starting point. The situation can be understood analytically as follows. 

Consider the case when $\a\sigma^2$ can always be neglected with respect to the other terms. That means that $\sigma $ must be 
sufficiently large for $\b\sigma^4$ to dominate, but cannot be too close to 1. Then 
\be
V\simeq \frac{e^{-2\b \sigma^4}}{\a}\left[1-\tilde g \sigma^4-4\b\sigma^4\left(1-\frac{\tilde g}{5}\sigma^4\right)\right]^2.
\ee
We would like to see if this potential doesn't have a maximum larger than the value at zero, i.e. $1/\a$. 

First, if we can neglect the $\tilde g \sigma^4/5$ term, the minimum is at 
\be
\sigma_{\rm min}\simeq \left[\frac{1}{\tilde g+4\b}\right]^{1/4}
\Rightarrow \b\sigma_{\rm min}^4\simeq \frac{\b}{\tilde g+4\b}.
\ee
It seems that simply imposing $\tilde g\ll \b$ would work, since then $\b\sigma_{\rm min}^4\simeq 1/4$, so the factor in front of the potential would be 
$e^{-2\b\sigma_{\rm min}^4}\simeq e^{-1/2}$ already. But a more precise calculation shows it still doesn't give the required result, 
since the extrema of the function
\be
F\equiv e^{-2\b\sigma^4}(1-4\b\sigma^4)^2\;,
\ee
defined by
\be
-8\b\sigma^3(1-4\b\sigma^4)e^{-2\b\sigma^4}[5-4\b\sigma^4]=0\;
\ee
are given by a minimum at $\b\sigma_{\rm min}^4=1/4$, and a maximum at $\b\sigma_{\rm max}^4=5/4$.  At the maximum we have 
\be
F=\frac{4^2}{e^{2.5}}\simeq \frac{16}{12.2}>1\;,
\ee
so the maximum is actually higher than the initial plateau.

We were able in fact to prove analytically that no value of $\b, \tilde g$ avoids this problem. Moreover, we were able to prove that 
neither does changing the $\tilde g \sigma^4/5$ power law into another power, nor changing $\b\sigma^4$ into any other function 
{\em analytic at zero} (Taylor expandable). In fact, the function 
replacing $\b\sigma^4$ needs to be non-analytic, more specifically nonperturbative, at $\sigma=0$. 

\subsection{Model with correct properties}

In order to obtain a model that doesn't have a maximum with higher value than the one at $\sigma=0$, we need to 
replace $\b\sigma^4$ with a function $f(\sigma)$ satisfying $\sigma f'(\sigma)\ll f(\sigma)-f(0)$ for $\sigma=\sigma_{min}\ll 1$ (that is, at the minimum 
{\em of the potential} V, the first correction in a would-be Taylor expansion is actually much smaller than the deviation from zero). 
Such a function is $e^{-\frac{c}{\sigma}}$. Yet we also still need the $e^{-2\b\sigma^4}$ factor to counteract the $e^{\a\sigma^2}$ 
factor at large $\sigma$,.
We will also see that the $\tilde g\sigma^4/5$ term is not useful anymore, so we will put $\tilde g=0$ later, but we will
just keep it for now for completeness. We then consider the superpotential
\be
W(\Phi)=e^{ib\Phi}e^{-\b_1\Phi^4-\b_2 e^{c/i\Phi}}\Phi\left(1-\frac{\tilde g}{5}\Phi^4\right).
\ee
The potential is then found to be 
\bea
V&\simeq&\frac{e^{-2b\sigma}e^{-2\b_1\sigma^4+\a\sigma^2-2\b_2e^{-c/\sigma}}}{\a}
\left\{\left[1-\tilde g \sigma^4+\a\sigma^2-\left(b\sigma+4\b_1\sigma^4+\frac{c}{\sigma}\b_2e^{-\frac{c}{\sigma}}\right)\left(1-\frac{\tilde g}{5}\sigma^4
\right)\right]^2\right.\cr
&&\left.-3\a\sigma^2\left(1-\frac{\tilde g}{5}\sigma^4\right)^2+...\right\}\;,\label{approxv}
\eea
and on the inflationary region we can approximate further,
\be
V\simeq\frac{e^{-2\b_1\sigma^4}}{\a}\left[1-4b\sigma +7b^2\sigma^2+0\cdot \a\sigma^2-2\left(\tilde g+4\b_1\right)\sigma^4+{\cal O}(b^3\sigma^3)\right].
\ee

We want to analyze the end of inflation and the presence of a maximum, so we ignore for the moment $\a\sigma^2$ and
$b\sigma$ in (\ref{approxv}), considering that we are in an intermediate region, after inflation, but before these terms become of order 1. 
We also consider that we are before the onset of the $\b_1\sigma^4$ terms, and we put $\tilde g=0$. 
Then we have 
\be
V\simeq \frac{e^{-2\b_2e^{-c/\sigma}}}{\a}\left[1-\frac{c}{\sigma}\b_2e^{-\frac{c}{\sigma}}\right]^2.\label{simplified}
\ee
To analyze this potential, consider the function 
\be
f(x)\equiv \frac{c}{x}e^{-\frac{c}{x}}\;
\ee
with derivative
\be
f'(x)=-\frac{c}{x^2}e^{-\frac{c}{x}}[1-\frac{c}{x}]
\ee
Then the extrema of $f(x)$ are at $c=x$, and since $f(0)=0$ and $f(\infty)=0$, at $x=c$ there is a maximum, with value $f(c)=1/e$. The condition 
that the simplified potential above (which is positive definite) is zero is $\b_2 f(x)=1$, or $f(x)=1/\b_2$. If $1/\b_2<1/e$, this equation will have exactly 
two solutions, one with $c/x>1$ and one with $c/x<1$. 

In this case the square factor in (\ref{simplified}), 
$(1-\b_2 f(x))^2$, will (for $\b_2>e$ by a margin) drop down to zero, then go to a maximum value of $(1-\b_2/e)^2$,
and then drop again. Then the value of the exponential prefactor of (\ref{simplified}) is $e^{-2\b_2e^{-c/x}}=e^{-2\b_2/e}$.
For the full potential at the maximum value, consider the function
\be
g(y)=e^{-2y}(1-y)^2
\ee
where $y=\b_2/e$, with derivative
\be
g'(y)=-2e^{-2y}(1-y)(2-y)\;,
\ee
which is positive between $y=1$ and $y=2$ and negative otherwise, meaning that $y=1$ is a minimum (as we knew), and $y=2$ is a maximum. 
But $g(0)=1,g(1)= 0$, whereas $g(2)=e^{-4}\ll 1$. That means that for the optimum value of $\b_2$ (and for any $c$), the value of the potential 
(\ref{simplified}) at the maximum cannot be larger than $e^{-4}\ll 1=V(0)$, which is what we wanted to obtain, so out model satisfies the 
desired conditions.

We still need a large enough value of $\b_2$, so that the suppression factor $e^{-2\b_2}$ overcomes any increase due to the 
$\b_1\sigma^4$ term (which is needed 
to stop at large $\sigma $ the $\a\sigma^2$ term in the exponential prefactor). We can choose for instance $\b_2\sim 10$ and $c\sim 1$, 
which means that 
the square in (\ref{simplified}) is maximum at $\sigma_{\rm max}=c\sim 1$. The two solutions for $V=0$ are obtained from the equation 
\be
\b_2 y=e^y
\ee
where now $y=c/x$, and are seen to be approximately $y\simeq 1/9$ and $y\simeq 4$ (since $e^4\simeq 54, \b_2 y\simeq 40$, $e^{1/9}\simeq 10/9$, 
$\b_2y\simeq 10/9$). This leads to the first minimum (at zero) being for $\sigma_{\rm min}\sim c/4\sim 1/4$ and the second minimum 
being for $\sigma_2\sim 9c\sim 9$.

We can also find when inflation ends as follows. The value of $\sigma$ there has to be less than 
$\sigma_{\rm min}\sim 1/4$. Consider $\sigma=1/8$. Then $\b_2 c/x e^{-c/x}\sim 
80 e^{-8}\simeq 80/2980\simeq 1/35$, whereas $\b_2e^{-c/x}\sim 10e^{-8}=10/2980\sim 1/300$, meaning the exponential prefactor $e^{-2\b_2e^{-c/x}}\sim 
e^{-1/150}$ is completely irrelevant. So at $\sigma=1/8$, the $\b_2$ term is completely irrelevant for the exponential, but it gives a $1/35$ correction to 
the 1 inside the square bracket, which is close to being relevant, whereas at $\sigma=1/4$ is already gives $V=0$. 
So the end of inflation is very shortly after $\sigma=1/8$, hence we can assume $\sigma_{\rm end}\simeq 1/8$. 

\subsection{K\"{a}hler potential modifications and the volume modulus}

We can now also consider a possible completion of the K\"{a}hler potential away from zero that does not affect the good characteristics of the 
potential. In particular, one motivation for the condition that $\sigma$ needs to be able to reach very large values is the case when $\sigma$ 
is a volume modulus for large extra dimensions. 

We have seen at the beginning of this section that exponential modifications to the K\"{a}hler potential do not give inflation, and also 
polynomial modifications can be problematic. One possible modification is an interpolation between the canonical $K=\a\bar \Phi\Phi$ 
at small values and the logarithmic $K=-\b \ln[-i(\bar\Phi-\Phi)]$ required for a volume modulus. Indeed, for a volume modulus, simple 
KK reduction on a compact space of the Einstein action leads to such a $K$ (more precisely, to the resulting scalar metric $g_{\bar\Phi\Phi}$), 
with $\b=3 $ for 6 extra dimensions, and $\b=2$ for 2 large extra dimensions (and 4 small ones). $\sigma$ is the volume of 4-cycles in the 
space. For more details, see \cite{Hinterbichler:2013we}.

Such an interpolation is naturally logarithmic: Indeed, quantum corrections to the logarithmic K\"{a}hler modulus appear usually inside the 
log, see for instance the review \cite{Quevedo:1996sv}. Since we want these modifications to be nonperturbative in nature, a natural K\"{a}hler 
potential that achieves this is 
\be
K=-2\ln\left[-i(\Phi-\bar\Phi)+\lambda e^{-\a\bar\Phi\Phi}\right].
\ee
This leads to a scalar metric for $\sigma$ of
\be
g_{\Phi\bar\Phi}=\d_{\bar\Phi}\d_\Phi K=\frac{2\left[1+\a\lambda^2e^{-2\a\sigma^2}-2\a^2\lambda\sigma^3e^{-\a
\sigma^2}\right]}{\left[2\sigma+\lambda e^{-\a\sigma^2}\right]^2}.
\ee
For the model from the above subsection, we obtain
\bea
K&=&-2\ln\left[2\sigma+\lambda e^{-\a\sigma^2}\right]\cr
W&=&i e^{-b\sigma}e^{-\b_1\sigma^4-\b_2e^{-c/\sigma}}\sigma\left(1-\frac{\tilde g}{5}\sigma^4\right)\cr
D_\sigma W&=&e^{-b\sigma}e^{-\b_1 \sigma^4-\b_2e^{-c/\sigma}}
\left[1-\tilde g \sigma^4-\left(b\sigma+4\b_1 \sigma^4+\frac{c}{\sigma}
\b_2e^{-\frac{c}{\sigma}}\right)\left(1-\frac{\tilde g}{5}\sigma^4\right)\right.\cr
&&\left.-2\frac{\left(1-\a\lambda\sigma e^{-\a\sigma^2}\right)\left(1-\frac{\tilde g}{5}\sigma^4\right)}{2+\frac{\lambda}{\sigma}e^{-\a\sigma^2}}\right].
\eea
The corrections due to $\lambda$ however (to deviation from the canonical form) can be ignored for the whole analysis of the inflationary 
part and the part just after the potential 

This model was considered in \cite{Hinterbichler:2013we}, and its consequences for inflation were discussed. However, it was seen that 
it leads to an $n_s-1$ that is in conflict with the latest Planck data: one obtains naturally a  value of the order of $10^{-5}$, as opposed to the 
observed $\sim 0.03$.

\section{Conclusions}

In this paper we have analyzed the possibility of having small field inflation in mimimal supergravity coupled with a single scalar superfield 
that contains the inflaton. While it is always possible to consider very contrived functional forms for the K\"{a}hler potential and superpotential, 
that combine exponentials, logs and powers in a nontrivial way, we have analyzed systematically simple possibilities for small values of the 
inflaton. 

Besides the previously identified class of models with a superpotential defined by a real function and a nontrivial K\"{a}hler potential (\ref{specialkahler}), 
that is difficult to obtain from a more fundamental construction (say in string theory), we have found only a very restricted set of possibilities. 
Only a canonical K\"{a}hler potential with corrections, $K=\bar\Phi\Phi+...$ with a Taylor expansion for the superpotential
starting at $W=A\Phi+...$ gives inflation. In order to avoid a very high fine tuning of the initial conditions, we obtain a constraint 
on the coefficients of the Taylor expansions. One possible solution for it would be $W=A\Phi e^{b\Phi^2}+...$. Another
solution has already been considered by 
Yzawa and Yanagida. 

If we also want to have a way to end inflation in a reheating or preheating phase, we need to have a steep drop in the potential, followed by a 
minimum. This further restricts the possibilities. The Izawa-Yanagida model (\ref{yy}) is one possibility. If one allows for a large value of the minimum, 
we can have other solutions, like $W=A\Phi e^{b\Phi^2}$. 

Finally, if we want to be able to reach large values of $\Phi$, like in the case that $\Phi$ is a volume modulus in a large extra dimensions scenario, 
there need to be no maximum larger than $V(\Phi=0)$ at larger $\Phi$, and we find that
all simple possibilities for $K$ and $W$ are excluded. Only very convoluted possibilies are allowed. We have found one such example, but besides being very 
difficult to see how it could be realized, it is already ruled out by the experimental values of $n_s-1$.

\section*{Acknowledgements}

We would like to thank Ricardo Landim for discussions.
HN's work is supported in part by CNPq grant 301709/2013-0 and FAPESP grant 2014/18634-9.
HB's work is supported by CAPES.

\bibliography{msugrainflation}
\bibliographystyle{utphys}

\end{document}